# The kinetic Monte Carlo Simulation scheme of the homoepicaxial growth of GaAs(001) for heterostructural growth on GaAs(001) substrate.

A. Ishii, M.Tsukao, N.Toda, and S.Oshima
Department of Applied Mathematics and Physics, Tottori University, Koyama, Tottori 680-8552, Japan,
ishii@damp.tottori-u.ac.jp

The simulation scheme for heterostructural growth of compound semiconductors is presented based on the kinetic Monte Carlo method. The sheme is designed as simple as possible in order to apply it for any heteroepitaxial growth on GaAs(001) substrate. The parameters used in the simulation are determined with the first-principles calculation to reproduce experimental RHEED intensity curves for homoepitaxial growth of GaAs(001).

## 1. Introduction

Nowadays, GaAs(001) surface is very important as a substrate for quantum dot(QD) growth with InAs. In order to use the quantum dot of InAs/GaAs(001) system for electronic and optical devices, it is significant to control the size and the position of the quantum dots in atomic scale. However, such control is still very difficult in real laboratories.

In order to control size and location of QD, the optimization of the heteroepitaxial growth condition is important. Thus, we should investigate the mechanism of QD formation. Though it is known that strain of InAs layers on GaAs substrate is siginificant to determine size of QD, we should know also the mechanism to determine the location of QD in atomic scale. Using the *in situ* fast scan STM[1], it begins to be possible to observe directly the growth process of QD formation in laboratories. Thus, in order to investigate the mechanism to determine the location of QD, the growth simulation for heteroepitaxial growth of InAs/GaAs(001) surface is required as a reference for experimental investigation.

We can investigate less than 1nm × 1nm area very accurately by using the first principles calculation. However, since the diffusion length of adatoms on the surface is longer than size of such area, the first principles calculation is not enough to investigate the epitaxial growth phenomena.

One of very powerful computational method to investigate the epitaxial growth is the kinetic Monte Carlo method (kMC)[2]. The kMC is a kind of Monte Carlo simulation to investigate the time development of a system having multiple dynamical processes using the random number. The kMC is well-known for studies on the growth of Si(001)[3] and GaAs(001)[4-8].

The purpose of the paper is to present the similar kMC simulation scheme for heteroepitaxial growth for III-V compound semiconductors.

## 2. kinetic Monte Carlo simulation

### 2.1 Model

For homoepaxial growth of GaAs(001), Itoh[8] presented very complicated simulation model based on the zincblende(001) structure. However, his approach is constraint to only β2(2×4) reconstructed structure of GaAs(001) surface. Such too much accurate approach would not be adequate for the purpose to simulate heteroepitaxial growth on GaAs(001) substrate. Even for homoepitaxial growth of GaAs(001), the well-known reconstructed structures such as β2(2×4), c(4×4)α, c(4×4)β, c(8×2), (4×6), etc. appear only after growth stopped. Since we do not know much about the detailed atomic processes of heteroepitaxial growth, it is better to assume in the simulation model that the lattice structure is just zincblende without any reconstruction. For adatom dynamics, anisotropy of diffusion and anisotropic incorporation to islands can be included in the simulation as a hopping barrier energy from a site to other site. The hopping barrier energies for Ga and As should be tuned to reproduce RHEED intensity oscillation observed in experiments.

Heteroepitaxy can be treated as a simulation with three atomic species, Ga, As and In. The strain effect between the substrate and adlayer can be included in the hopping barrier energy. In this article, we present a simulation model for heteroepitaxial growth on GaAs(001) substrate with a set of adjusted parameters for Ga and As which can reproduce surface step density or RHEED intensity oscillation curves for homoepitaxial growth of GaAs(001).

### 2.2 Model for homoepitaxial growth of GaAs(001)

In order to find the best condition of the epitaxial growth, the dynamical feature of Ga and As atoms on the surface is required. The dynamical feature can be simulated by using the kinetic Monte Carlo (kMC) method. In the kMC method, we can simulate rather easily the time evolution of the growth as stacking of atomic processes. The key parameters in the kMC method is the hopping barrier energies of each atoms from a site to a neighbor site on the surface.

The algorithm of the kMC method was presented by Bortz et.al.[9] and it was applied to the MBE by Maksym[2]. The advantage of this algorithm is that we can simulate time-dependent phenomena consisting of several time-dependent events occuring in parallel. The key parameters for the kMC simulation is hopping barrier energies for atoms. Under the thermodynamic equilibrium, the migration rate R is

$$R = R_0 \exp\left(-\frac{E_b}{kT}\right)$$

where $R_0$ is the prefactor, $k$ is Boltzmann constant and $T$ is the substrate temperature. $E_b$ is the barrier energy for adatom hop to a neighbor site. $E_b$ should be determined to reproduce experiments. Typically, the prefactor R0 is taken to be nearly a inverse number of the lattice vibration frequency.

The rate of the arrival of atoms on the surface is also determined. The summation of the rate over all migrations and arrivals gives the rate of the event. For each steps, the event occured inteh step is chosen using the random number.

In this study, the model of the kMC simulation is based on the work of Kawamura[3] who employ the realistic simulation scheme with diamond structure (001) surface instead of the simple SOS model. Here we extend his model for zincblende structure (001) surface. The extension is mainly to use multi atomic species.

The barrier energy is defined by the enviroment of the atom, so that the barrier energies for isolated adatom on the terrace is different from the barrier energy of the atom incorporated at the step edge. For the case of the homoepitxial growth of GaAs(001), the migration barrier energies for Ga and As are assumed to be defined from the number of the first and second nearest atoms for each atoms in order to compare them with the first-principle calculation in near future. Thus, the barrier energy is defined as follows

$$E_b = n_1 E_1 + n_2 E_2$$

where $E_1$ and $E_2$ are the binding energy for the first and the second nearest atoms, and $n_1$ and $n_2$ are the number of the first and the second nearest occupied atomic sites during growth.

Since we should treat Ga and As individually in the simulation, we consider the barrier energy both for Ga and As atoms.

$$E_{Ga} = n_{1Ga} E_{Ga-As} + n_{2Ga} E_{Ga-Ga}$$

$$E_{As} = n_{1As} E_{Ga-As} + n_{2As} E_{As-As}$$

$E_{Ga-As}$ is the Ga-As effective binding energy for hopping. The barrier energies due to the second nearest neighbour are different for Ga and As: $E_{Ga-Ga}$ is the Ga barrier energy due to the second nearest neighbour Ga atoms and $E_{As-As}$ is the As one. It should be noted that the each energies $E_{Ga-As}$, $E_{Ga-Ga}$ and $E_{As-As}$ do not correspond directly to the bond breaking energy, because the migrating atom is still adsorbed on the surface even at the highest barrier energy position.

For GaAs(001) surface, especially for β2(2×4) surface, anisotropy of islands in growth is well-known, so that we should include anisotropy of movement and incorporation of adatoms. The anisotropy observed in the morphology is caused by the two reasons; dimer-dimer correlation and anisotropy of hopping ratio itself. The dimer-dimer correlation can be included as the same way as that of the model of Kawamura[3] for Si(001) surface: the dimer formation energy $E_{2D}$ and the dimer-dimer correlation energy along the dimer row direction, $E_{2DR}$ are included. The intrinsic anisotropy for hopping of adatom itself on a terrace is also included as $E_{Ga\_anisotropy}$ and $E_{As\_anisotropy}$ where the barrier energy is determined for each hoppong path. It was not included in the model of Kawamura[3]. In our kMC model, the rate of migration is determined independently for each migration path in contrast to the model of Kawamura[3]. Thus, the hopping barrier energy for Ga and As is assumed to be as follows.

$$E_{Ga} = n_{1Ga}E_{Ga-As} + n_{2Ga}E_{Ga-Ga} + n_{2DGa} E_{2DGa} + n_{2DRGa} E_{2DRGa} + E_{Ga\_anisotropy}$$

$$E_{As} = n_{1As}E_{Ga-As} + n_{2As}E_{As-As} + n_{2DAs} E_{2DAs} + n_{2DRAs} E_{2DRAs} + E_{As\_anisotropy}$$

where $n2_{DGa}$ and $n2_{DAs}$ are number of atoms which can form a dimer with the adatom along the dimer direction. Thus, $n2_{DGa}$ and $n2_{DAs}$ are 0 or 1. $n2_{DRGa}$ and $n2_{DRAs}$ are number of dimer nearest to the adatom along the dimer row direction. $E_{Ga\_anisotropy}$ and $E_{As\_anisotropy}$ are the anisotropy of substrate itself.

**2.3 Parameter adjustment for kMC model**

For the surface of GaAs(001)- β2(2×4), anisotropy of the islands during growth is known to be very strong. The islands elongate along the dimer direction in contrast to Si(001) surface where islands elongate along the dimer row direction with similar dimer structure. Because Ga layers and As layers are treated separately as the zincblende (001) surface structure, mixed dimer structure of c(4×4) structure cannot be modeled in our simulation. In order to reproduce the anisotropy of islands growth elongated along the dimer direction for GaAs(001)- β2(2×4) surface, we set to be $E_{2DAs}>0$ and $E_{2DRAs}=0$. Using such assumption, only the anisotropy of islands are included and the detailed dimer formation and the atomic trough appeared in β2(2×4) structure is excluded. Therefore, we can apply our simuation model for As-dmer-c(4×4) structure also.

In our simulation model, we assume that there are very local and very temporal Ga-rich region on the surface where As adatoms diffuse and adsorb. Since the local and temporal structure during growth is not required thermodynamically stable because of thermally non-equilibrium condition, quasi-stable structure can appear. We assume that the local and temporal Ga-rich region can be consisted with two or three Ga dimers. We assume that the hopping motion of As on such Ga-rich region can be emulated with the hopping motion of As adatom on the Ga-terminated GaAs(001)- β2(4×2) surface which is considered to have similar anisotropy as GaAs(001)- β2(2×4).

In kMC simulation, we must determine the barrier energy $E_b$ by a trial and error method or obtain $E_b$ from other methods like the first-principle calculation. In this paper, we determine the barrier energy by a trial and error to adjust the surface step density curve with the RHEED intensity oscillation curve measured in the experiment. In the recent study[10], the surface step density oscillation curve corresponds nearly to the RHEED intensity oscillation curve, though we should be careful to the diffraction condition which can affect the relative phase of the RHEED oscillation. $E_{Ga\_anisotropy}$ and

$E_{As\_anisotropy}$ are determined from the first principles calculation directly that $E_{Ga\_anisotropy}$ is 0.3eV and $E_{As\_anisotropy}$ is 0.2eV for dimer row direction.

## 2.4 Model for heteroepitaxial growth of InAs/GaAs(001)

To apply the kMC simulation for heteroepitaxial growth, we should treat multi atomic species in the calculation. Since the crystal structure itself is same for InAs and GaAs, the atomic site definition is the same zincblende (001) surface with slightly different lattice constant.
Namely, the hopping barrier energies for In, Ga, and As adatoms are defined as follows;

$E_{In} = n_{1In}E_{In-As} + n_{2In-In}E_{In-In} + n_{2In-Ga}E_{In-Ga} + n_{2DIn} E_{2DIn} + n_{2DRIn} E_{2DRIn} + E_{In\_anisotropy}$

$E_{Ga} = n_{1Ga}E_{Ga-As} + n_{2Ga-Ga}E_{Ga-Ga} + n_{2Ga-In}E_{Ga-In} + n_{2DGa} E_{2DGa} + n_{2DRGa} E_{2DRGa} + E_{Ga\_anisotropy}$

$E_{As} = n_{1As-In}E_{As-In} + n_{1As-Ga}E_{As-Ga} + n_{2As}E_{As-As} + n_{2DAs} E_{2DAs} + n_{2DRAs} E_{2DRAs} + E_{As\_anisotropy}$

Since the substrate GaAs(001) surface has anisotropy, we assume that $E_{2DAs}=0$ and $E_{2DRAs}>0$. Since InAs wetting layer on GaAs(001) has no anisotropy for island growth, we can assume that $E_{2DAs}=0$ and $E_{2DRAs}=0$.
In the real simulation, the diffusion of In adatoms on InAs wetting layer on GaAs(001) and on intrinsic InAs(001) surface are predicted to be different using the first principles calculation[11]. Thus, the hopping barrier energy for In adatom should be calculated with different parameters for In adaom on the wetting layer and on the thick InAs islnad (quantun dot).

## 3. Results and Discussion

### 3.1 Comparison with experiment

We adjust the parameters for Ga and As as a base of the heteroepitaxial growth simulation. In fig.1, we show the adjusted negative surface step density curves for homoepitaxial growth of two vicinal surfaces of GaAs(001) - β2(2×4), the vicinal toward [110] and [1$\bar{1}$0] direction or A-surface and B-surface where the substrate temperature is 556°C and the supplied beam intensities are set to be 0.4ML/sec for Ga and 2.0ML/sec for As assuming As$_2$. The corresponding experiment is RHEED intensity oscillation curuves measured by Shitara et al.[12] where the beam intensity is 0.4ML/sec for Ga and 2.0ML/sec for As using As$_2$ beam source.
The typical feature of the difference between the A-surface and the B-surface is that the growth is more step-flow like for the B-surface, especially at 556°C. In other words, the oscillation is larger for A-surface. The other important feature is that the oscillation behavior appears up to three periods. The oscillation strength degreases with increasing the time of growth and the oscillation seems to be very weak at the fourth period.
The curves showed in fig.1 is the best adjusted curves with the RHEED intensity oscillation curves of ref.12. The parameters used in fig.1 is shown in table 1. This set of parameters can be used also in the heteroepitaxial growth based on the GaAs(001) - β2(2×4) substrate surface.

### 3.2 Comparison with the first principles calculations

In our kinetic Monte Carlo scheme, the adjusted hopping barrier energy for Ga on truncated As-terminated GaAs(001) surface is calculated using table 1 as $2E_{Ga-As} + 4E_{Ga-Ga} = 0.928eV$. Similary, the adjusted barrier energy for As adatom on truncated Ga-terminted surface is 0.728eV. It agrees with the first principles calculations[7, 13,14] that As adatoms are more mobile than Ga adatoms.

In the first principles calculation, the GaAs(001) surface is not the truncated surface but reconstructed surface. For example, under slight As-rich condition, the most stable reconstructed surface structure is known to be GaAs(001)-β2(2×4) surface where the easiest diffusion path is in the trough of this reconstructed structure along the As dimer direction. The diffusion barrier energy along the path is 1.2eV in the former first principles calculation[15]. The hopping barrier energy for Ga adatom on the trough of GaAs(001)-β2(2×4) surface is calculated as $2E_{Ga-As}$ with two As atom forming As dimer in the bottom of the trough, $4E_{Ga-Ga}$ with four Ga atom under the As dimer of the third layer and $2E_{Ga-Ga}$ with Ga atom forming the wall of the trough with same height as that of the Ga adatom in the trough. The calculated value for the Ga adatom in the trough is 1.22eV. Similarly, we obtain 0.92eV for As adatom on the trough of Ga-terminated GaAs(001)-β2(4×2) surface. The first principles calculation shows us 1.1eV[7]. Therefore, the parameter set of table 1 shows us similar diffusion barrier energy compared with the first principles calculation.

The other example is the barrier energy for As adatom on GaAs(001)-ζ(4×2) surface which was formerly accepted reconstructed surface structure of GaAs(001) in Ga-rich condition. The barrier energy for As adatom on this surface using the first principles calculation[14] is 0.5eV along $[1\bar{1}0]$ direction. The most stable position for As adatom is AA site in ref.14. For the AA site, the As adatom has one first nearest neighbor Ga atom, two second nearest neighbor As atoms of the topmost layer and one second nearest neighbor As atom connected with the first nearest Ga. Thus, using the parameter value in table 1, we obtain the barrier energy for As as 0.461eV. This value is also very similar to the value obtained by using the first principles calculation.

### 3.3 Extension to heteroepitaxial growth

After we checked the diffusion barrier parameters for Ga and As for homoepitaxial growth of GaAs(001) surface, the simulation for heteroepitaxial growth on GaAs(001) substrate will be possible. For example for InAs/GaAs(001), we should determine $E_{In-As}$, $E_{In-In}$, $E_{In-Ga}$, $E_{2DIn}$, $E_{2DRIn}$, and $E_{In\_anisotropy}$. According to the discussion of the former section, the first principles calculation of diffusion barrier energy for In adatom on GaAs(001) surface[11] will be very helpful to determine the parameters.

### 4. Conclusion

The kinetic Monte Carlo smulation shceme for heteroepitaxial growth of InAs/GaAs(001) is presented. The parameters for diffusion of Ga and As adatoms on the surface are set to be nearly equal to the results of the first principles calculation. The parameters are checked that the RHEED intensity curves for homoepitaxial growth for two vicinal surfaces of GaAs(001) can be reproduced in the simulation. Adding diffusion parameters for Indium adatom, we will be able to perform heteroepitaxial growth of InAs/GaAs(001) surface.

**Acknowledgement**
The work is supported partially by the Grant-in-Aid of the Ministry of Education, Science and Cultures of Japan. The authors are grateful for Profs. T.Kawamura and T.Aisaka for valuable discussions.

|    | parameter in kMC | adjusted value [eV] |
|----|------------------|---------------------|
| Ga | $E_{Ga-As}$      | 0.170               |
|    | $E_{Ga-Ga}$      | 0.147               |
|    | $E_{2DGa}$       | 0.555               |
| As | $E_{As-Ga}$      | 0.170               |
|    | $E_{As-As}$      | 0.097               |
|    | $E_{2DAs}$       | 0.490               |

Table 1
The adjusted parameters for homoepitaxial growth of GaAs(001) used in the simulation of Figure 1.

Figure 1

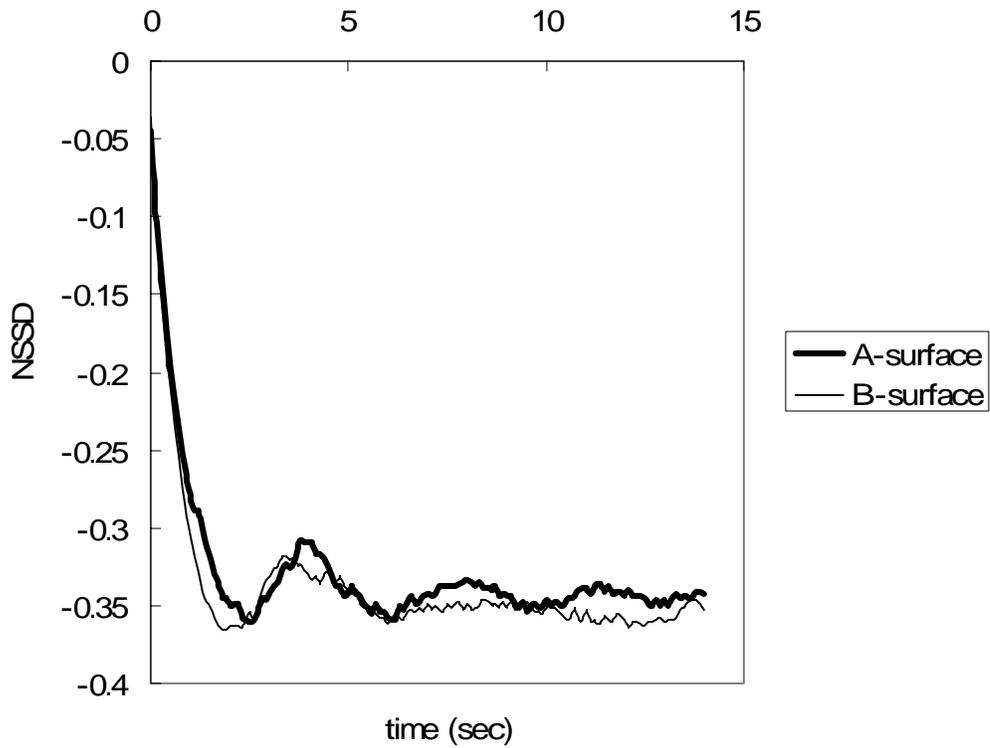

Figure 1
The calculated surface step density for homoepitxial growth of GaAs(001). A-surface and B-surface correspond to the vicinal surface of 2° toward [110] and [1$\bar{1}$0] direction. The temperature is set to be 556°C.